\documentclass[sigconf,balance=false]{acmart}
\AtBeginDocument{%
  }

\acmDOI{10.1145/3858063.3858801}

\setcopyright{none}
\copyrightyear{2026}
\acmYear{2026}
\acmConference[ICAIF '26]
  {7th ACM International Conference on AI in Finance}
  {November 14--17, 2026}
  {Milan, Italy}

\acmBooktitle{7th ACM International Conference on AI in Finance
  (ICAIF '26), November 14--17, 2026, Milan, Italy}

\acmISBN{979-8-4007-3087-0/2026/11}

\renewcommand\footnotetextcopyrightpermission[1]{}
\begin{document}

\title[MLP-Initialized RL for SPX Short-Straddle Exit Timing]{MLP-Initialized Reinforcement Learning for Exit Timing in SPX Short-Straddle Portfolios}
\titlenote{Full version of the paper accepted at ACM ICAIF 2026 (Milan, November 14--17, 2026).}
\author{Jianning Wang}
\orcid{0009-0006-5578-9936}
\affiliation{%
  \institution{School of Physical and Mathematical Sciences, Nanyang Technological University}
  \country{Singapore}
}
\email{JWANG106@e.ntu.edu.sg}

\author{Robert Ślepaczuk}
\correspondingauthor
\orcid{0000-0001-5227-2014}
\affiliation{%
  \institution{University of Warsaw, Department of Quantitative Finance and Machine Learning}
  \city{Warsaw}
  \country{Poland}
}
\email{rslepaczuk@wne.uw.edu.pl}

\author{Paweł Sakowski}
\orcid{0000-0003-3384-3795}
\affiliation{%
  \institution{University of Warsaw, Department of Quantitative Finance and Machine Learning}
  \city{Warsaw}
  \country{Poland}
}
\email{p.sakowski@uw.edu.pl}

\renewcommand{\shortauthors}{Wang et al.}

\begin{abstract}
This paper studies exit timing for SPX short-straddle portfolios using a two-stage supervised and reinforcement-learning framework. Historical episodes first generate ex-post optimal hold/exit labels for training a multilayer perceptron (MLP) using only contemporaneously available state variables at deployment. The trained MLP then initializes reinforcement-learning actors that are fine-tuned with a risk-aware trading reward. Out-of-sample evaluation over 2024--2026 shows that the supervised MLP substantially outperforms passive holding and representative fixed-time and profit-and-loss-based exit rules. Across seven independent seeds, MLP-initialized Actor--Critic averages 37.54\% cumulative return and a Sharpe ratio of 1.702, compared with seven-seed averages of 35.57\% and 1.603 for the supervised MLP. The incremental gain is economically positive, while matched-episode tests provide mixed evidence. Actor--Critic trained from scratch averages only 0.18\% return. These findings show that oracle-guided supervised learning provides a stable foundation for reward-based policy fine-tuning in a sparse and path-dependent option-exit problem.
\end{abstract}





\maketitle

\section{Introduction}

Short-volatility strategies seek to earn the premium embedded in option prices but remain exposed to large market moves and volatility expansions. A short straddle, which sells a call and put with the same strike and expiration, is a canonical example linked to the variance risk premium \cite{carr2009variance,bollerslev2009expected}. Its asymmetric loss profile makes the timing of position closure central to practical performance.

This paper studies when an existing SPX short straddle should be closed before expiration. At each decision time, a policy observes contemporaneous market, quote, and position variables and chooses whether to hold or exit. We propose a two-stage framework. First, ex-post optimal exits from historical training episodes generate labels for a supervised multilayer perceptron (MLP). Second, the trained MLP initializes reinforcement-learning actors that are fine-tuned using a risk-aware trading reward. Future information is used only for offline training-label construction and never as an out-of-sample policy input.

All policies are evaluated on the same locked 2024--2026 out-of-sample episodes under common bid--ask execution, commissions, settlement, position sizing, and portfolio accounting. The supervised MLP substantially outperforms hold-to-expiry and representative fixed-time and profit-and-loss-based rules. Across seven seeds, MLP-initialized Actor--Critic raises average cumulative return from 35.57\% to 37.54\% and average Sharpe ratio from 1.603 to 1.702. Matched-episode tests provide mixed evidence for the increment, so we interpret it as a descriptive average improvement. AWR and AWAC broadly preserve supervised performance, whereas raw Actor--Critic averages only 0.18\% return, highlighting the importance of supervised initialization.

The contribution applies supervised initialization and reward-based fine-tuning to option exit timing. This paper makes three contributions. First, it formulates SPX short-straddle exit timing as a sequential learning problem under realistic execution and portfolio accounting. Second, it introduces an oracle-guided supervised policy that initializes reward-based reinforcement-learning fine-tuning. Third, it provides multi-seed evidence against passive, mechanical, and raw-RL controls, together with reward-shaping, synthetic-fill, and equal-notional robustness checks.

\section{Background and Related Work}
\subsection{Short Straddles and Exit Timing}

A straddle is an options strategy that combines a call and a put option with the same underlying asset, strike price, and expiration date~\cite{oic_options_strategies_quick_guide}. This paper focuses on short straddles on SPX index options. In a short straddle, the trader sells both the call and the put and receives the initial option premium. The strategy benefits when the underlying index remains close to the strike and the option time value decays before expiration.

For a short straddle with strike $K$, call premium $C_0$, and put premium $P_0$, the expiration profit before transaction costs can be written as
\begin{equation}
  \label{eq:short_straddle_payoff}
  \Pi_T = C_0 + P_0 - |S_T-K|.
\end{equation}

This study examines the voluntary exit timing problem for existing short-straddle positions before expiration. Since SPX index options are European-style and cash-settled \cite{cboeSPXSpecifications}, terminal payoff is determined by cash settlement rather than physical delivery. Holding too long may expose the portfolio to tail losses, while exiting too early may sacrifice remaining premium. Exit timing is therefore central to converting the short-volatility premium into a more stable trading strategy.

Simple time-based and profit-and-loss-based exits provide important benchmarks for evaluating a learned exit policy. Fixed-horizon exits test whether performance can be explained by shortening exposure mechanically, while profit targets and stop losses test whether a constant threshold is sufficient to control losses or harvest gains. Stop-loss rules are widely used in practice, but their effectiveness depends on return dynamics, persistence, and transaction costs rather than following mechanically from loss limitation alone \cite{kaminski2014stoploss}.

A fair comparison requires these rules to be evaluated on the same short-straddle opportunities and under the same execution and portfolio-accounting assumptions as the learned policy. The present study therefore uses hold-to-expiry, fixed-time, profit-target and stop-loss, and combined exits as progressively more active non-learning benchmarks. These comparisons distinguish state-dependent exit timing from the passive variance-risk premium and from improvements that could be reproduced by a constant calendar or profit-and-loss rule.

\subsection{Variance Risk Premium and Volatility Selling}

Short-option strategies are often motivated by the variance risk premium. In equity index markets, implied volatility in option prices tends to exceed subsequent realized volatility, so option sellers may earn compensation for bearing volatility and tail risk \cite{carr2009variance,bollerslev2009expected}. A short straddle is a natural volatility-selling strategy because it receives premium upfront and profits when the realized movement of the underlying index remains limited.

This premium comes with substantial downside risk. Short-volatility strategies can generate frequent small profits but suffer large losses when the underlying index moves sharply or volatility expands. Existing literature links option returns and volatility risk premia to compensation for systematic volatility and downside risk \cite{coval2001expected,bakshi2003delta}. Therefore, average return alone is not sufficient for evaluating a short-straddle strategy.

Accordingly, this paper evaluates each exit policy using portfolio return, Sharpe ratio, and maximum drawdown. The purpose of exit timing is to retain the premium-harvesting benefit of short straddles while reducing exposure to large adverse moves.

\subsection{Machine Learning for Trading Decisions}

Machine learning provides a natural way to model trading decisions from market and contract-level state variables. In this paper, the exit decision is formulated as a binary classification problem. At each decision time $t$, the model observes a state vector $s_t$ that summarizes the current short-straddle position and market condition, and outputs the probability of closing the position:
\begin{displaymath}
  f_{\theta}(s_t) = \Pr(a_t = \mathrm{exit} \mid s_t).
\end{displaymath}
The state vector includes variables such as moneyness, time-to-maturity, option quote features, spread-related measures, volatility-risk-premium signals, and position holding-time features.

The training labels are generated from ex-post optimal exit decisions within historical contract episodes. For each historical short-straddle episode, the best feasible exit time is computed using the realized future path under the same transaction-cost and settlement assumptions used in the backtest. States before the selected exit time are labeled as hold, while the selected exit state and subsequent states are labeled as exit. This transforms a noisy sequential decision problem into a supervised classification problem.

This construction is related to imitation learning and learning from demonstrations, in which a policy is trained to reproduce actions supplied by an expert or oracle \cite{argall2009survey,ross2011dagger}. The setting here is specifically offline oracle-guided classification: the oracle labels a fixed historical dataset, and the MLP is not allowed to query the oracle interactively or observe future outcomes at deployment. Unlike interactive imitation-learning methods such as DAgger, the proposed procedure does not aggregate new oracle labels from states visited by the learned policy.

Future information is used only offline to construct training labels and to calculate a non-tradable diagnostic ceiling. At test time, the MLP observes only contemporaneously available state variables and does not observe future prices or future option values. The supervised MLP is a tradable policy, whereas the ex-post optimal exit rule is neither tradable nor included as a competing deployable strategy.

\subsection{Reinforcement Learning for Trading}

Reinforcement learning (RL) studies sequential decision problems in which an agent interacts with an environment and learns from rewards \cite{sutton2018reinforcement}. The exit-timing problem can be represented as a Markov decision process with state $s_t$, action $a_t$, reward $r_t$, and transition dynamics. In this paper, the action space is binary:
\begin{displaymath}
  a_t \in \{\mathrm{hold}, \mathrm{exit}\}.
\end{displaymath}
A policy $\pi_\theta(a_t \mid s_t)$ maps the current trading state to a distribution over actions. The objective is to maximize expected discounted reward:
\begin{equation}
  \max_{\theta}
  \mathbb{E}_{\pi_\theta}
  \left[
    \sum_{t=0}^{T} \gamma^t r_t
  \right].
\end{equation}
Here, $\gamma \in [0,1]$ is the discount factor.

In trading applications, RL is attractive because it can optimize sequential objectives that include transaction costs, delayed rewards, and path-dependent risk. However, direct RL training from random initialization is often unstable in financial markets, where rewards are noisy, nonstationary, and dominated by rare adverse events. This is especially relevant for short-option strategies, which may generate frequent small gains but occasional large losses. Therefore, this paper uses RL primarily as a second-stage optimizer initialized from a supervised MLP exit policy, rather than relying only on raw RL from scratch.

\section{Data and Trading Environment}

\subsection{Data Source and Coverage}

The option data consist of licensed Cboe SPX index-option quotations accessed by the research team through an institutional data agreement held by one of the co-authors. The dataset contains 30-minute observations from 2016 through 2026 for near-the-money SPX options with moneyness ratios between 0.95 and 1.05 and remaining maturities from 0 to 21 trading days. The available fields include call and put bid and ask quotations, strike price, expiration, implied volatility, option Greeks, and the contemporaneous SPX index level.

The 30-minute sampling scheme produces up to 13 intraday decision times per trading day. Data from 2016--2023 are used for label construction, supervised learning, and reinforcement-learning fine-tuning, while observations from 2024--2026 form the locked out-of-sample evaluation period. The S\&P 500 series displayed as a market reference in the NAV figure is obtained separately from Yahoo Finance using the \texttt{yfinance} package \cite{yahooFinanceGSPC,yfinance}; it is not used to construct the SPX option trades or train the exit policies.

\subsection{SPX Short-Straddle Episodes}

The raw option panel is filtered before episode construction. Observations with non-finite bid, ask, implied-volatility, or underlying-index values are removed. The underlying index level must be positive, option bids must be non-negative, and observations for which the ask is below the bid are excluded. Duplicate contract-time observations are removed, and call and put records are matched by expiration, strike, and quote timestamp. A straddle observation is retained only when both option legs are available at the same timestamp. These filters prevent non-finite, negative, or internally inconsistent quotations from directly determining the reported entry and exit fills.

Across the complete episode set, synthetic continuation bars account for 5.40\% of observations. Their shares are 5.69\% in the 2016--2023 training period and 5.03\% in the 2024--2026 out-of-sample period. A detailed audit of episode coverage, strategy exits, and conservative synthetic-fill assumptions is reported in Appendix~\ref{app:synthetic_fill_audit}.

Matched call--put quotes are grouped by expiration and strike. For each contract, an episode begins at the earliest quote satisfying 18--22 trading days to maturity and moneyness $K/S_t\in[0.975,1.025]$. For each start date, one candidate is selected by minimizing, in order, the distance from 20 trading days to maturity, the distance from at-the-money, the relative straddle spread, and the quote time. If a portfolio slot is available, the evaluator attempts to sell the selected call and put at their combined bid at the episode's first quote. If the capital and short-margin checks cannot support one contract, entry is deferred and retried at subsequent real quotes; if no slot is available, that day's candidate is skipped. Entry selection is therefore mechanical and common to all strategies, while the learned policies control only subsequent hold/exit decisions. After entry, strike and expiration remain fixed until policy exit or cash settlement. SPX options are European-style and cash-settled \cite{cboeSPXSpecifications}.

At each decision time $t$, the policy observes the current state of the open position, including the underlying index level, moneyness, remaining time-to-maturity, quote and spread information, volatility-related signals, holding time, and unrealized profit and loss. The available action is binary:
\begin{displaymath}
  a_t \in \{\mathrm{hold}, \mathrm{exit}\}.
\end{displaymath}
Discretionary exit is disabled during the common 26-bar minimum holding period. If the policy chooses hold, the position remains open. If it chooses exit, the short straddle is closed and the episode ends. If no exit occurs before expiration, the position is cash-settled at terminal intrinsic value.

\subsection{Portfolio Construction}

The backtest is conducted at the portfolio level rather than as isolated single-contract episodes. The portfolio uses 20 reusable capital slots to represent overlapping TTM-20 short-straddle positions. Since a new candidate trade is introduced each trading day and each contract has an approximately 20-day target maturity, the 20 slots form a rolling portfolio structure.

The portfolio starts with \$3 million divided equally across the 20 slots. Each daily candidate is assigned to the free slot with the earliest availability date, with ties broken by slot identifier; it is skipped if none is free. Each slot can hold at most one active short-straddle position. The same trained exit policy is applied to every active slot. When a position exits or expires, the corresponding slot is reused for the next entry assigned to that slot. Thus, the 20-slot structure should be interpreted as portfolio capital allocation rather than 20 independently trained agents.

\subsection{State Variables}

All tradable policies use the same observation vector constructed by the trading environment at each decision time. In the final experiments, the environment uses the position-augmented state specification, so the state vector has dimension 17. The state is written as
\begin{equation}
  s_t =
  \left[
    x_t,\; p_t
  \right].
\end{equation}
Here, $x_t$ contains market and quote variables, and $p_t$ contains position-specific variables. In the implementation, ratio features are computed with standard numerical safeguards and the final observation is clipped to the range $[-20,20]$.

The liquidity-adjusted score is $\mathrm{VRP}_t/\mathrm{SpreadRel}_t$, with zero assigned when the relative spread is zero. The base market and quote component is
\begin{displaymath}
\begin{aligned}
x_t =
\big[
&\tau_t/260,\;
\mathrm{tod}_t,\;
\mathrm{VRP}_t/10,\;
\mathrm{spreadrel}_t/\bar{s}_{\mathrm{rel}},\\
&\mathrm{score}_t/\bar{q},\;
\Delta \mathrm{RV13}_t/\sigma_{\Delta \mathrm{RV13}},\\
&\mathbf{1}\{\mathrm{quote\ available}\},\;
\mathbf{1}\{\mathrm{synthetic\ bar}\},\\
&\mathrm{timevalue}_t/\mathrm{mid}_t,\;
\mathrm{pos}_t,\;
\mathrm{PnL}_t
\big].
\end{aligned}
\end{displaymath}
Here $\tau_t$ is the remaining number of bars to expiration, $\mathrm{tod}_t$ is the time-of-day variable, $\mathrm{VRP}_t$ is the volatility-risk-premium proxy, $\mathrm{spreadrel}_t$ is the relative bid--ask spread, and $\mathrm{score}_t$ is the liquidity-adjusted VRP score. The constants $\bar{s}_{\mathrm{rel}}$, $\bar{q}$, and $\sigma_{\Delta \mathrm{RV13}}$ are computed from the training sample and used to normalize the relative spread, score, and realized-volatility-change feature. The fixed divisor of 10 rescales the VRP proxy to a magnitude comparable with the other numerical state inputs without changing its sign or ordering. The clipping interval $[-20,20]$ limits the influence of extreme ratios and numerical outliers on neural-network optimization. Both preprocessing conventions are fixed before out-of-sample evaluation and applied identically to all policies.

The position-specific component is
\begin{displaymath}
\begin{aligned}
p_t =
\big[
&h_t/260,\;
\tau^{\mathrm{entry}}/260,\;
\tau_t/\tau^{\mathrm{entry}},\\
&V_t/V^{\mathrm{entry}},\;
(S_t-S^{\mathrm{entry}})/S^{\mathrm{entry}},\\
&m_t-m^{\mathrm{entry}}
\big].
\end{aligned}
\end{displaymath}
Here $h_t$ is the number of bars held since entry, $\tau^{\mathrm{entry}}$ is the time-to-maturity at entry, $V_t$ is the current position value, $V^{\mathrm{entry}}$ is the entry value, $S_t$ is the current index level, and $m_t=K/S_t$ is moneyness. The variable $\mathrm{pos}_t$ equals $+1$ for an active short straddle and $0$ for a flat slot. The unrealized PnL feature is set to zero when no position is active.

\begin{table}[t]
  \centering
  \caption{State variables used by the exit policies.}
  \label{tab:state_variables}
  \resizebox{\linewidth}{!}{%
  \begin{tabular}{lll}
    \toprule
    Index & Variable & Description \\
    \midrule
    1 & $\tau_t/260$ & Remaining bars to expiration \\
    2 & $\mathrm{tod}_t$ & Time of day \\
    3 & $\mathrm{VRP}_t/10$ & VRP proxy \\
    4 & $\mathrm{spreadrel}_t/\bar{s}_{\mathrm{rel}}$ & Relative spread \\
    5 & $\mathrm{score}_t/\bar{q}$ & Liquidity-adjusted VRP score \\
    6 & $\Delta \mathrm{RV13}_t/\sigma_{\Delta \mathrm{RV13}}$ & Realized-volatility change \\
    7 & Quote flag & Quote availability indicator \\
    8 & Synthetic flag & Synthetic-bar indicator \\
    9 & $\mathrm{timevalue}_t/\mathrm{mid}_t$ & Remaining time-value ratio \\
    10 & $\mathrm{pos}_t$ & Position indicator \\
    11 & $\mathrm{PnL}_t$ & Current normalized PnL \\
    12 & $h_t/260$ & Holding time \\
    13 & $\tau^{\mathrm{entry}}/260$ & Entry time-to-maturity \\
    14 & $\tau_t/\tau^{\mathrm{entry}}$ & Fraction of entry maturity remaining \\
    15 & $V_t/V^{\mathrm{entry}}$ & Current value relative to entry value \\
    16 & $(S_t-S^{\mathrm{entry}})/S^{\mathrm{entry}}$ & Underlying return since entry \\
    17 & $m_t-m^{\mathrm{entry}}$ & Moneyness change since entry \\
    \bottomrule
  \end{tabular}%
  }
\end{table}

The state design reflects the economic structure of the exit-timing problem. Time-to-maturity and time of day describe where the contract is in its lifecycle. Quote and spread variables capture the cost of closing the position, which is important because option bid--ask spreads vary with contract characteristics and trading activity \cite{george1993bidask,wei2010trading}. Volatility and VRP-related features reflect the short-volatility exposure of the strategy and the role of variance risk premia in option markets \cite{carr2009variance,bollerslev2009expected}. Position variables describe how far the active trade has moved from entry in both option-value and underlying-index terms.

Only information available at the decision time is included in $s_t$. Future prices and future option values are used only offline to construct supervised exit labels and diagnostic benchmarks. This separation prevents look-ahead bias in out-of-sample evaluation.

\subsection{Transaction Costs and Settlement}

Transaction costs and bid--ask execution are included throughout the portfolio simulation. The implementation uses the same execution rule for all tradable strategies. When a short straddle is opened, the strategy sells the straddle at the bid price:
\begin{displaymath}
  F_t^{\mathrm{entry}} = B_t^{\mathrm{straddle}}.
\end{displaymath}
Here, $B_t^{\mathrm{straddle}}$ is the combined bid price of the call and put legs. When an active short straddle is closed before expiration, the strategy buys back the straddle at the ask price:
\begin{displaymath}
  F_t^{\mathrm{exit}} = A_t^{\mathrm{straddle}}.
\end{displaymath}
If a real quote is not available during a simulated continuation segment, the exit cost is replaced by the simulated short-exit cost used by the environment. Synthetic continuation bars are explicitly flagged in the state and are never presented as observed market quotations. Appendix~\ref{app:synthetic_fill_audit} reports their frequency and evaluates more conservative synthetic half-spread assumptions.

For a position with $n$ contracts and option multiplier $M$, the gross dollar profit from closing a short straddle before expiration is
\begin{displaymath}
  \mathrm{PnL}^{\mathrm{gross}}_t
  =
  nM
  \left(
    F_0^{\mathrm{entry}} - F_t^{\mathrm{exit}}
  \right).
\end{displaymath}
The simulation also charges a fixed commission per option contract. Since a straddle has two option legs, the commission paid at each entry or exit event is
\begin{displaymath}
  \mathrm{Comm}
  =
  2cn.
\end{displaymath}
Here, $c$ is the commission per option contract. In the experiments, $c=\$1.50$. Therefore, a round-trip short-straddle trade incurs commissions on both entry and exit. Including these costs is important because trading frictions can materially affect option strategy performance \cite{leland1985option,demsetz1968cost,george1993bidask,wei2010trading}.

If a position is held to expiration, the terminal settlement value is the intrinsic value of the straddle:
\begin{displaymath}
  I_T
  =
  |S_T-K|.
\end{displaymath}
The gross dollar profit at expiration is therefore
\begin{equation}
  \mathrm{PnL}^{\mathrm{gross}}_T
  =
  nM
  \left(
    F_0^{\mathrm{entry}} - I_T
  \right).
\end{equation}
This settlement rule is consistent with the European-style, cash-settled nature of SPX index options \cite{cboeSPXSpecifications}. The same execution, settlement, and accounting conventions are used for the hold-to-expiry benchmark, the rule-based exit policies, the supervised MLP policy, and all reinforcement-learning policies.

\subsection{Evaluation Metrics}

All strategies are evaluated using the same out-of-sample portfolio simulation. The performance panel reports cumulative return, annualized compounded return (ARC), annualized standard deviation (ASD), maximum drawdown magnitude (MD), maximum loss duration (MLD), three reward-to-risk measures, Sharpe ratio, Calmar ratio, hit rate, and number of trades. Mean episode profit and loss is reported in the text and in the robustness tables. ARC annualizes portfolio growth over the evaluation horizon, while ASD measures annualized return volatility. MD measures the largest peak-to-trough portfolio loss, and MLD measures the longest period during which portfolio NAV remains below its previous high. Unless noted otherwise, NAV-based metrics use the common real 30-minute quote grid and calendar horizon $T \approx 2.42$ years. The principal AC+MLP panel retains the archived per-run evaluation summaries used in the conference camera-ready; its risk and annualized-return statistics are not re-estimated from the common-grid plotting series.

The three additional reward-to-risk measures are defined as
\begin{displaymath}
  \mathrm{IR}^{*}=\frac{\mathrm{ARC}}{\mathrm{ASD}},
  \qquad
  \mathrm{IR}^{**}=\frac{\mathrm{ARC}}{\mathrm{MD}},
  \qquad
  \mathrm{IR}^{***}=\frac{\mathrm{ARC}}{\mathrm{MLD}}.
\end{displaymath}
Under these definitions, $\mathrm{IR}^{**}$ is numerically equal to the Calmar ratio. Hit rate is the fraction of entered short-straddle episodes with positive net dollar PnL, and the trade count is the number of entered straddles. Cumulative return is retained for economic interpretation, while ARC provides the horizon-consistent annualized return measure required for comparison with annualized risk statistics. Maximum drawdown and loss duration are particularly relevant for short-volatility strategies because losses may be both severe and persistent \cite{rockafellar2000optimization,rockafellar2002conditional}. Sharpe ratios use 30-minute marked-to-market NAV returns, annualized by $\sqrt{252\times13}$, without subtracting a separate risk-free rate. Episode tests match start date, expiration, strike, episode identifier, and daily start index. For the main multi-seed comparison, each run is first filtered to actual entries; PnL is then averaged across available seeds within each policy and episode before joining the policies. Unmatched episodes remain in portfolio metrics but are excluded from paired tests.

The deployable benchmark strategies include hold-to-expiry, three representative rule-based exit policies, and raw reinforcement learning from scratch. The rule-based policies consist of a fixed-time exit, a profit-target and stop-loss rule, and a combined rule. The proposed tradable models include the supervised MLP exit policy and MLP-initialized reinforcement-learning variants. Separately, the ex-post best-exit oracle is reported as a non-tradable diagnostic ceiling. Because it observes each episode's realized future path, it is excluded from the main comparison of deployable strategies.

\section{Methodology}

\subsection{Stage 1: Supervised MLP Exit Policy}

The first stage converts the exit-timing problem into a supervised binary classification problem. For each historical short-straddle episode $i$, let $t=0,\ldots,T_i$ denote the sequence of decision times after entry. The position is first opened at $t=0$, and the subsequent states are used to construct an exit-label dataset.

\begin{figure}[t]
  \centering
  \includegraphics[width=\linewidth]{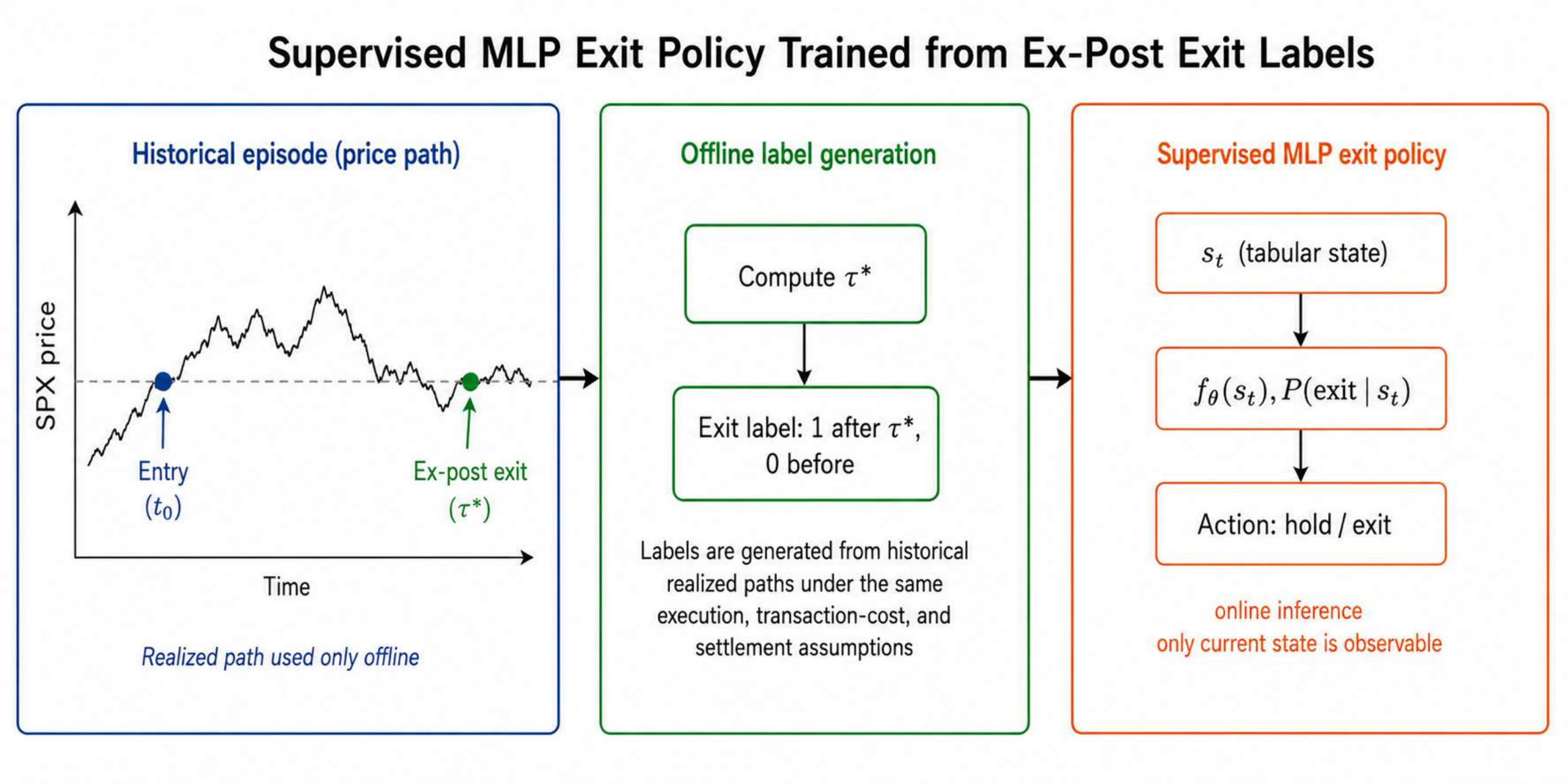}
  \caption{Supervised MLP exit-policy training from ex-post exit labels.}
  \Description{Historical short-straddle episodes are used offline to compute ex-post optimal exit times. These labels train an MLP classifier that maps current state variables to an exit probability.}
  \label{fig:mlp_labeling_flow}
\end{figure}

Let $F_{i,0}^{\mathrm{entry}}$ denote the entry fill price of the short straddle. For each feasible exit time $t$, define $F_{i,t}^{\mathrm{exit}}$ as the cost of closing the short straddle. If the contract has reached expiration, the exit value is the terminal intrinsic value,
\begin{displaymath}
  F_{i,t}^{\mathrm{exit}} = |S_{i,t}-K_i|.
\end{displaymath}
Otherwise, the exit value is computed from the available quote using the same bid--ask execution rule used in the portfolio backtest.

The realized dollar profit from exiting episode $i$ at time $t$ is
\begin{equation}
  \Pi_{i,t}^{\mathrm{exit}}
  =
  M\left(F_{i,0}^{\mathrm{entry}} - F_{i,t}^{\mathrm{exit}}\right)
  -
  4c.
\end{equation}
Here, $M$ is the option contract multiplier and $c$ is the commission per option contract. The factor $4c$ accounts for two option legs at entry and two option legs at exit.

A minimum holding constraint is imposed before an exit label can be assigned. Let $h$ denote the minimum holding period. In all reported experiments, $h=26$ intraday bars, corresponding to approximately two trading days at 13 observations per trading day. The ex-post optimal exit time is then defined as
\begin{equation}
  \tau_i^\star
  =
  \arg\max_{t \in \{h,\ldots,T_i\}}
  \Pi_{i,t}^{\mathrm{exit}}.
\end{equation}
This step uses the realized historical path only offline. It is not available to the trading policy during out-of-sample evaluation.

The supervised label is defined as
\begin{equation}
  y_{i,t}
  =
  \mathbf{1}\{t \ge \tau_i^\star\}.
\end{equation}

This label design teaches the classifier not only to exit at the ex-post optimal time, but also to keep recommending exit if the position remains open after that point.

Given the observed state vector $s_{i,t}$, the MLP outputs an exit logit
\begin{displaymath}
  z_{i,t} = f_\theta(\tilde{s}_{i,t}).
\end{displaymath}
Here, the input is standardized using training-sample mean and standard deviation,
\begin{displaymath}
  \tilde{s}_{i,t}
  =
  \frac{s_{i,t}-\mu_{\mathrm{train}}}
       {\sigma_{\mathrm{train}}}.
\end{displaymath}
The predicted exit probability is
\begin{displaymath}
  \hat p_{i,t}
  =
  \sigma(z_{i,t})
  =
  \Pr(y_{i,t}=1\mid s_{i,t}).
\end{displaymath}
Here, $\sigma(\cdot)$ is the logistic sigmoid function.

The network is a fully connected MLP with two hidden layers and ReLU activations:
\begin{equation}
  f_\theta(\tilde{s})
  =
  W_3
  \phi\!\left(
    W_2
    \phi(W_1\tilde{s}+b_1)
    +b_2
  \right)
  + b_3.
\end{equation}
Here, $\phi(x)=\max(x,0)$.

Because exit labels are sparse relative to hold labels, the model is trained using weighted binary cross-entropy:
\begin{equation}
  \mathcal{L}_{\mathrm{MLP}}
  =
  -
  \sum_{i,t}
  \left[
    \omega_+ y_{i,t}\log \hat p_{i,t}
    +
    (1-y_{i,t})\log(1-\hat p_{i,t})
  \right].
\end{equation}
Here, the positive-class weight is
\begin{displaymath}
  \omega_+
  =
  \frac{N_{\mathrm{hold}}}{N_{\mathrm{exit}}}.
\end{displaymath}
This weighting matches the implementation and compensates for the imbalance between hold and exit observations.

Figure~\ref{fig:mlp_labeling_flow} summarizes this supervised training pipeline. Historical realized paths are used offline to generate ex-post exit labels, while the trained MLP policy uses only contemporaneously available state variables during out-of-sample trading.

\subsection{Stage 2: Reinforcement Learning Fine-Tuning}

The two stages optimize related but distinct objectives. Stage 1 imitates the ex-post exit that maximizes realized episode PnL after transaction costs, providing a stable policy initialization. Stage 2 fine-tunes this policy using a sequential reward that additionally reflects mark-to-market changes, volatility-premium exposure, liquidity costs, and downside risk. The supervised policy is therefore an informed starting point rather than an oracle-equivalent solution. The second stage refines the supervised exit policy using reinforcement learning. After Stage 1, the trained MLP provides a strong initialization for the RL actor. In other words, the same neural network parameterization is reused as the policy network, but its weights are no longer optimized against hindsight labels. Instead, they are further updated to improve trading performance in the short-straddle environment \cite{sutton2018reinforcement,konda2000actorcritic}.

\begin{figure}[t]
  \centering
  \includegraphics[width=\linewidth]{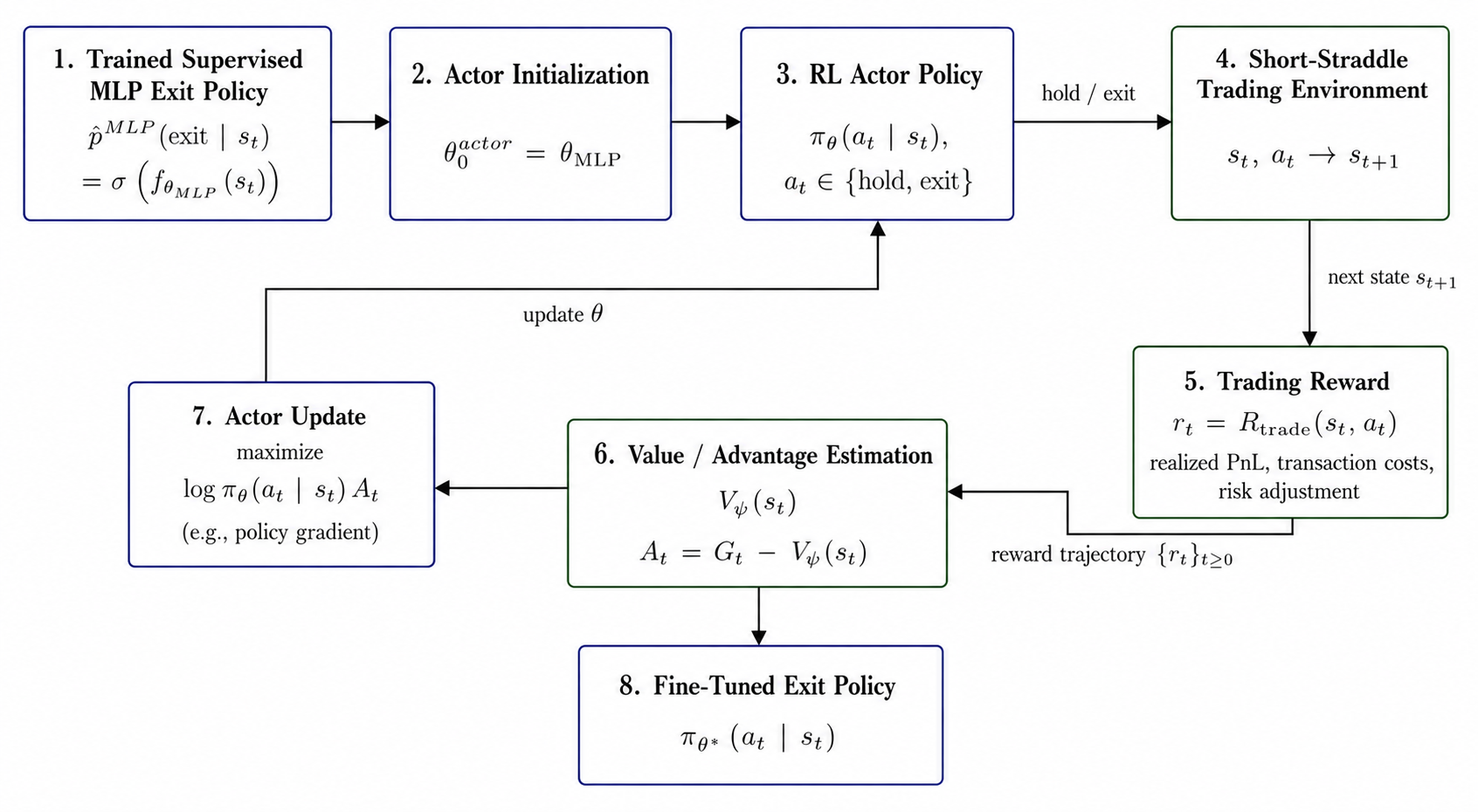}
  \caption{MLP-initialized reinforcement learning fine-tuning.}
  \Description{The supervised MLP initializes the RL actor, which interacts with the trading environment and is updated using realized trading rewards.}
  \label{fig:rl_finetuning_process}
\end{figure}

Figure~\ref{fig:rl_finetuning_process} summarizes the fine-tuning pipeline. The supervised MLP initializes the RL actor, which then interacts with the trading environment and is updated using realized trading rewards.

At each decision time $t$, the agent observes the current state $s_t$ and selects an action
\begin{displaymath}
  a_t \in \{\mathrm{hold}, \mathrm{exit}\}.
\end{displaymath}
The action is executed in the trading environment, which returns a trading reward $r_t$ computed from the realized outcome of the chosen action. In the final experiments, the same minimum holding constraint used in the supervised stage is retained, with $h=26$ bars.

The actor defines a policy $\pi_\theta(a_t \mid s_t)$ over the binary action space, and the supervised MLP is used to initialize the actor parameters,
\begin{displaymath}
  \theta_0^{\mathrm{actor}} = \theta^{\mathrm{MLP}}.
\end{displaymath}
This initialization allows reinforcement learning to start from a policy that already captures the main exit patterns learned from historical data, rather than from a random policy.

The reinforcement learning objective is to maximize the expected discounted return,
\begin{equation}
  J(\theta)
  =
  \mathbb{E}_{\pi_\theta}
  \left[
    \sum_{t=0}^{T_i} \gamma^t r_t
  \right].
\end{equation}
During training, the actor parameters are updated using trading rewards instead of supervised labels. In actor-critic style updates, a value function $V_\psi(s_t)$ is learned in parallel, and the advantage signal
\begin{displaymath}
  A_t = G_t - V_\psi(s_t).
\end{displaymath}
is used to improve the policy. Here $G_t$ denotes the realized return-to-go from time $t$.

The policy-gradient update can be written as
\begin{equation}
  \nabla_\theta J(\theta)
  \approx
  \sum_{i,t}
  \nabla_\theta \log \pi_\theta(a_{i,t} \mid s_{i,t}) A_{i,t}.
\end{equation}
In the advantage-weighted variants used in the experiments, actions with higher advantage receive larger update weight, which encourages the policy to favor exits that improve trading performance while preserving the stabilizing effect of the supervised initialization \cite{peng2019advantage,nair2020awac}.

The key idea is that Stage 1 provides an informed policy initialization, while Stage 2 adapts that policy to the actual trading objective through reward-driven optimization.

\subsection{RL Training Variants and Raw RL From Scratch}

This subsection describes the reinforcement learning variants used in the empirical study. All RL models share the same trading environment, state representation, action space, and transaction-cost assumptions. This follows the standard Markov decision process formulation, in which a policy maps observed states into actions and is optimized using realized rewards \cite{sutton2018reinforcement}. In this paper, the reinforcement-learning variants differ only in initialization and policy-update objective.

The environment uses the shaped trading reward implemented in the final RL training runs. Let $V_t^{\mathrm{before}}$ denote the value of the active position before the action is applied, and let $V_t^{\mathrm{after}}$ denote the value after the action. For an active position with entry value $V^{\mathrm{entry}}$, the normalized value change is
\begin{displaymath}
  \Delta v_t
  =
  \frac{
    \left(
      V_t^{\mathrm{before}} - V_t^{\mathrm{after}}
    \right)
    \mathrm{pos}_{t-}
  }{
    V^{\mathrm{entry}}
  }.
\end{displaymath}
Here, $\mathrm{pos}_{t-}$ is the position before the action. Under the environment convention, a short straddle has $\mathrm{pos}_{t-}=+1$.

The final RL experiments use the environment's shaped reward mode. The reward is computed by a single environment reward function. For an active short-straddle position, let $\Delta v_t$ denote the normalized change in position value, $\mathrm{PnL}_t$ denote the normalized realized profit and loss, $\mathrm{TC}_t$ denote the spread-based transaction-cost penalty, and $\tau_t$ denote the remaining time to maturity measured in intraday bars. The shaped reward is
\begin{equation}
  \label{eq:hold_reward}
\begin{aligned}
  r_t^{\mathrm{hold}}
  =
  &\Delta v_t
  +
  \beta_{\mathrm{VRP}}
  \mathrm{pos}_{t-}
  \mathrm{VRP}_t
  -
  \mathrm{TC}_t \\
  &-
  \lambda C_t
  \mathbf{1}\{\mathrm{pos}_{t-}=+1\}
  -
  \mu
  \mathbf{1}\{\tau_t \le 65,\;\mathrm{pos}_{t-}\ne 0\}.
\end{aligned}
\end{equation}
The environment's shaped reward is then
\begin{equation}
  r_t
  =
  \begin{cases}
  \mathrm{PnL}_t - \mathrm{TC}_t,
  & a_t=\mathrm{exit} \text{ or } t \text{ is terminal},\\
  r_t^{\mathrm{hold}},
  & \text{otherwise}.
  \end{cases}
\end{equation}
The first branch is used when the position is closed or settled, so the learning signal is the realized normalized PnL net of transaction cost. The second branch is used while the position remains active. It combines mark-to-market improvement, a VRP holding term, transaction-cost control, a CVaR-based short-volatility risk penalty, and a late-holding penalty.

The transaction-cost penalty is
\begin{equation}
  \label{eq:reward_tc}
  \mathrm{TC}_t
  =
  \alpha\,
  \mathrm{spreadrel}_t
  \mathbf{1}\{a_t\in\mathcal{A}_{\mathrm{trade}}\}.
\end{equation}
In the exit-only setting, this penalty is applied when the policy executes a closing trade. To define the tail-risk signal, let
\[
R_u^{\mathrm{short}}
=
\frac{V_{u-1}^{\mathrm{mid}}-V_u^{\mathrm{mid}}}
     {V_{u-1}^{\mathrm{mid}}}
\]
denote the one-bar return of a short straddle, where $V_u^{\mathrm{mid}}$ is its mid-quote value. Let $q_{0.05,t}$ be the empirical 5\% quantile of the available short-straddle returns in the trailing 260-bar window $\mathcal{W}_t$. The CVaR signal is the positive magnitude of the average left-tail return:
\begin{equation}
  \label{eq:cvar_signal}
  C_t
  =
  -
  \frac{
    \sum_{u\in\mathcal{W}_t}
    R_u^{\mathrm{short}}
    \mathbf{1}\{R_u^{\mathrm{short}}\le q_{0.05,t}\}
  }{
    \sum_{u\in\mathcal{W}_t}
    \mathbf{1}\{R_u^{\mathrm{short}}\le q_{0.05,t}\}
  }.
\end{equation}
The contract-level estimate requires at least 30 historical observations. When fewer observations are available, the contemporaneous 5\% left-tail CVaR estimate of the underlying index return is used as a fallback. Thus, larger $C_t$ represents greater estimated downside exposure for the active short straddle.

In the final RL training runs, the coefficients are set to $\alpha=1.5$, $\beta_{\mathrm{VRP}}=0.5$, $\lambda=0.2$, and $\mu=0.05$. This reward design encourages the agent to hold short straddles when the volatility premium is attractive, but penalizes liquidity costs, tail-risk exposure, and late-stage holding risk. At exit or expiration, the reward uses normalized PnL net of spread costs; fixed dollar commissions are applied separately in the final portfolio evaluation.

The coefficients were carried forward from a preliminary grid search trained on 2016--2021 data and ranked on 500 evenly sampled 2022--2023 validation episodes. Candidate configurations used $\alpha\in\{1.5,2.0\}$, $\lambda\in\{0.1,0.2\}$, $\beta_{\mathrm{VRP}}\in\{0.1,0.25,0.5\}$, and $\mu=0.05$, and were ranked by mean normalized final PnL minus one half of the 5\% PnL-CVaR loss magnitude and $0.001$ times mean trade count. The selected coefficients were then fixed across the final comparisons without using 2024--2026 performance. A multi-seed component ablation separately removes CVaR, VRP, or both shaping terms.

The first distinction across RL variants is initialization. In the proposed MLP-initialized setting, the actor starts from the supervised MLP parameters:
\begin{equation}
  \theta_0^{\mathrm{actor}} = \theta^{\mathrm{MLP}}.
\end{equation}
This uses the supervised exit model as a warm start for policy optimization. Such initialization is useful because neural policies can be sensitive to random initialization and optimization instability, especially in nonlinear financial control problems \cite{glorot2010understanding,goodfellow2016deep,buehler2019deep}. In the raw RL baseline, the actor is initialized randomly:
\begin{displaymath}
  \theta_0^{\mathrm{raw}} \sim \mathcal{I}.
\end{displaymath}
Here, $\mathcal{I}$ denotes the default neural-network initialization. The raw RL baseline therefore tests whether reinforcement learning alone is sufficient for the exit-timing task.

The standard actor-critic variant follows the policy-gradient framework with a learned value baseline \cite{williams1992simple,sutton2018reinforcement,konda2000actorcritic,schulman2016gae}. For each trajectory, define the realized return-to-go as
\begin{equation}
  \label{eq:return_to_go}
  G_t
  =
  \sum_{k=t}^{T}
  \gamma^{k-t}r_k.
\end{equation}
The critic estimates the state value $V_\psi(s_t)$ and is trained by minimizing
\begin{equation}
  \label{eq:value_loss}
  \mathcal{L}_V
  =
  \sum_t
  \left(
    V_\psi(s_t)-G_t
  \right)^2.
\end{equation}
The advantage estimate is
\begin{equation}
  \label{eq:advantage}
  A_t
  =
  G_t-V_\psi(s_t).
\end{equation}
The actor is then updated to increase the probability of actions with positive advantage:
\begin{equation}
  \mathcal{L}_{\mathrm{AC}}
  =
  -
  \sum_t
  \log \pi_\theta(a_t \mid s_t) A_t.
\end{equation}
Thus, if an exit or hold action produces better realized return than expected, the policy becomes more likely to choose that action in similar states. This advantage-based update is particularly useful in trading problems because it separates realized reward from the state-dependent baseline value \cite{konda2000actorcritic,moody2001learning,wen2021reinforcement}.

The AWR variant uses advantage-weighted regression \cite{peng2019advantage}. Instead of applying the advantage linearly, AWR converts the advantage into a positive sample weight:
\begin{equation}
  \label{eq:awr_weight}
  w_t
  =
  \exp\left(\frac{A_t}{\beta}\right).
\end{equation}
Here, $\beta>0$ is a temperature parameter. The actor objective is
\begin{equation}
  \label{eq:awr_objective}
  \mathcal{L}_{\mathrm{AWR}}
  =
  -
  \sum_t
  w_t
  \log \pi_\theta(a_t \mid s_t).
\end{equation}
This objective can be interpreted as weighted supervised learning on actions sampled from the trading trajectories, where actions with higher estimated advantage receive larger training weight. This makes AWR conceptually aligned with the two-stage design in this paper: the policy is initialized by supervised learning and then refined by advantage-weighted updates.

The AWAC-style variant uses the same return-to-go and state-value critic as the AWR variant rather than learning a separate action-value critic. It uses the advantage estimate in Equation~\ref{eq:advantage}, where $G_t$ is the realized discounted return-to-go and $V_\psi$ is trained using Equation~\ref{eq:value_loss}. The AWAC-style weight applies a non-negative advantage filter:
\begin{equation}
  \label{eq:awac_weight}
  w_t^{\mathrm{AWAC}}
  =
  \exp\left(
    \frac{\max\{A_t,0\}}{\beta}
  \right).
\end{equation}
The weight is clipped at the same maximum value used in the implementation. The actor objective is
\begin{equation}
  \label{eq:awac_objective}
  \mathcal{L}_{\mathrm{AWAC}}
  =
  -
  \sum_t
  w_t^{\mathrm{AWAC}}
  \log \pi_\theta(a_t \mid s_t).
\end{equation}
Thus, actions whose realized return exceeds the learned state-value baseline receive increased weight, while negative-advantage actions retain unit weight rather than being exponentially downweighted. We refer to this implementation as an AWAC-style variant because it uses non-negative advantage-weighted actor updates, although it does not train a separate $Q_\psi(s_t,a_t)$ critic.

Table~\ref{tab:rl_training_variants} summarizes the training variants used in the methodology.

\begin{table}[t]
  \centering
  \caption{Reinforcement learning training variants.}
  \label{tab:rl_training_variants}
  \begin{tabular}{lll}
    \toprule
    Variant & Initialization & Policy update \\
    \midrule
    Raw RL & Random & RL reward only \\
    Actor-Critic & MLP & Policy gradient with advantage \\
    AWR & MLP & Advantage-weighted regression \\
    AWAC-style & MLP & Non-negative advantage weights \\
    \bottomrule
  \end{tabular}
\end{table}

The purpose of including raw RL from scratch is methodological rather than merely empirical. If the raw RL policy performs poorly while MLP-initialized variants improve, then the results support the paper's central claim: reinforcement learning is most effective in this problem when used as a second-stage optimizer initialized from a supervised exit policy, rather than as a standalone learner. This is consistent with prior work showing that reinforcement learning can be useful in option and hedging applications, but that financial decision problems remain sensitive to reward design, transaction costs, and training stability \cite{buehler2019deep,cao2020deep,du2020deep,halperin2019qlbs,wen2021reinforcement}.

\section{Experimental Design}
\subsection{Train/Test Split}

The external evaluation uses a chronological pre-test/test split; the supervised training procedure also uses an internal random split of pre-test state rows. The held-out test period respects the time ordering of option markets. All model training and parameter estimation are performed using data before the out-of-sample evaluation period, while the final portfolio performance is measured on a later test period.

The locked out-of-sample period is January 2, 2024 to June 5, 2026 (approximately 2.42 years). Supervised label construction and final RL training use 2016--2023 data. The supervised implementation randomly reserves 20\% of pre-test state rows for internal validation, estimates normalization on the remaining training rows, and saves the final epoch. This internal split can place correlated states from the same episode on both sides and is not an independent chronological validation experiment. The seven supervised sensitivity models use eight epochs; the common supervised checkpoint used to initialize the RL actors was trained for ten epochs.

Reward coefficients were inherited from a preliminary search trained on 2016--2021 episodes and ranked on 500 evenly sampled 2022--2023 validation episodes. This preliminary reward search is distinct from supervised internal validation and does not imply that all final optimizer settings were selected by that search. No model hyperparameters or checkpoints were selected using the locked 2024--2026 test results. Future paths are used only offline for pre-test label construction and after decisions for out-of-sample performance measurement.

The same principle is applied to feature standardization. The mean and standard deviation used to normalize the MLP inputs are estimated from the training sample only. These training-sample statistics are then applied unchanged to validation and out-of-sample observations. This prevents the model from using information about the distribution of future test data.

All strategies are evaluated under the same out-of-sample portfolio simulator. The hold-to-expiry benchmark, representative rule-based exit policies, raw RL baseline, supervised MLP policy, and MLP-initialized RL variants are applied to the same test-period episode pool and portfolio construction rules. This ensures that performance differences reflect the exit policy and its resulting capital reuse rather than differences in data, execution assumptions, or evaluation protocol.

\subsection{Evaluation Protocol and Main Comparisons}

All exit strategies are evaluated using the same out-of-sample portfolio simulator over 2024--2026. Each strategy is applied to the same pool of SPX short-straddle episodes and uses the same daily candidate-selection rule, initial capital, 20-slot portfolio structure, stress-budget position sizing, bid--ask execution, commissions, margin requirements, and maintenance-liquidation rule. The component that differs across strategies is the exit decision. Because earlier exits release portfolio slots sooner, the resulting number of accepted episodes and the rate of capital reuse are endogenous consequences of each exit policy.

In the main specification, cumulative portfolio return is size-weighted and incorporates dynamic compounding within each slot. Under stress-budget sizing, the number of contracts assigned to a new position is computed from the slot's current NAV. A profitable slot can therefore support a larger subsequent position, while losses reduce its later position size. The reported main return consequently reflects both exit-policy decisions and the portfolio consequences of reinvesting gains under the common sizing rule. To separate exit-policy performance from this dynamic compounding channel, an equal-notional robustness specification fixes each entered position at one short-straddle contract.

The evaluation includes five groups of exit policies. The first is a hold-to-expiry benchmark, in which each short straddle remains open until cash settlement. This benchmark measures the passive short-volatility premium without discretionary exit timing.

The second group consists of three representative rule-based exit benchmarks. The fixed-time rule closes a position after 65 intraday bars, corresponding to approximately five trading days. The threshold rule exits when profit reaches 25\% of the entry premium or when loss reaches 50\% of the entry premium. The combined rule exits when either the fixed holding horizon or one of the profit-and-loss thresholds is reached. All three rules retain the same 26-bar minimum holding constraint used by the learned exit policies. These transparent heuristic baselines are not optimized on test-period performance. They test whether the learned models provide value beyond simple calendar-based and profit-and-loss-based exit heuristics.

The third policy is a raw Actor-Critic model trained from random initialization. This model serves as a control for whether reinforcement learning alone can discover a useful exit policy in this sparse and path-dependent trading environment.

The fourth policy is the supervised MLP exit model. It is trained from ex-post exit labels constructed only from training-period episodes and is applied out of sample using contemporaneously available state variables. This comparison measures how much exit-timing value can be captured through supervised learning.

The fifth group comprises Actor-Critic, AWR, and AWAC policies initialized from the MLP. These models initialize the actor from the supervised MLP and subsequently fine-tune the policy using the trading reward. This group tests whether reward-based policy improvement provides incremental value beyond the supervised exit model.

In addition to these five deployable groups, the evaluation reports an ex-post best-exit oracle as a diagnostic ceiling. For each out-of-sample episode, the oracle uses the complete realized future path to select the most profitable feasible exit after the same 26-bar minimum holding period. It uses the same bid--ask execution, commissions, settlement, position sizing, and portfolio simulator as the tradable policies. However, because its decision requires future information, it is reported separately and is not included in the main tradable-strategy table or NAV comparison.

\begin{figure}[t]
  \centering
  \includegraphics[width=\linewidth]{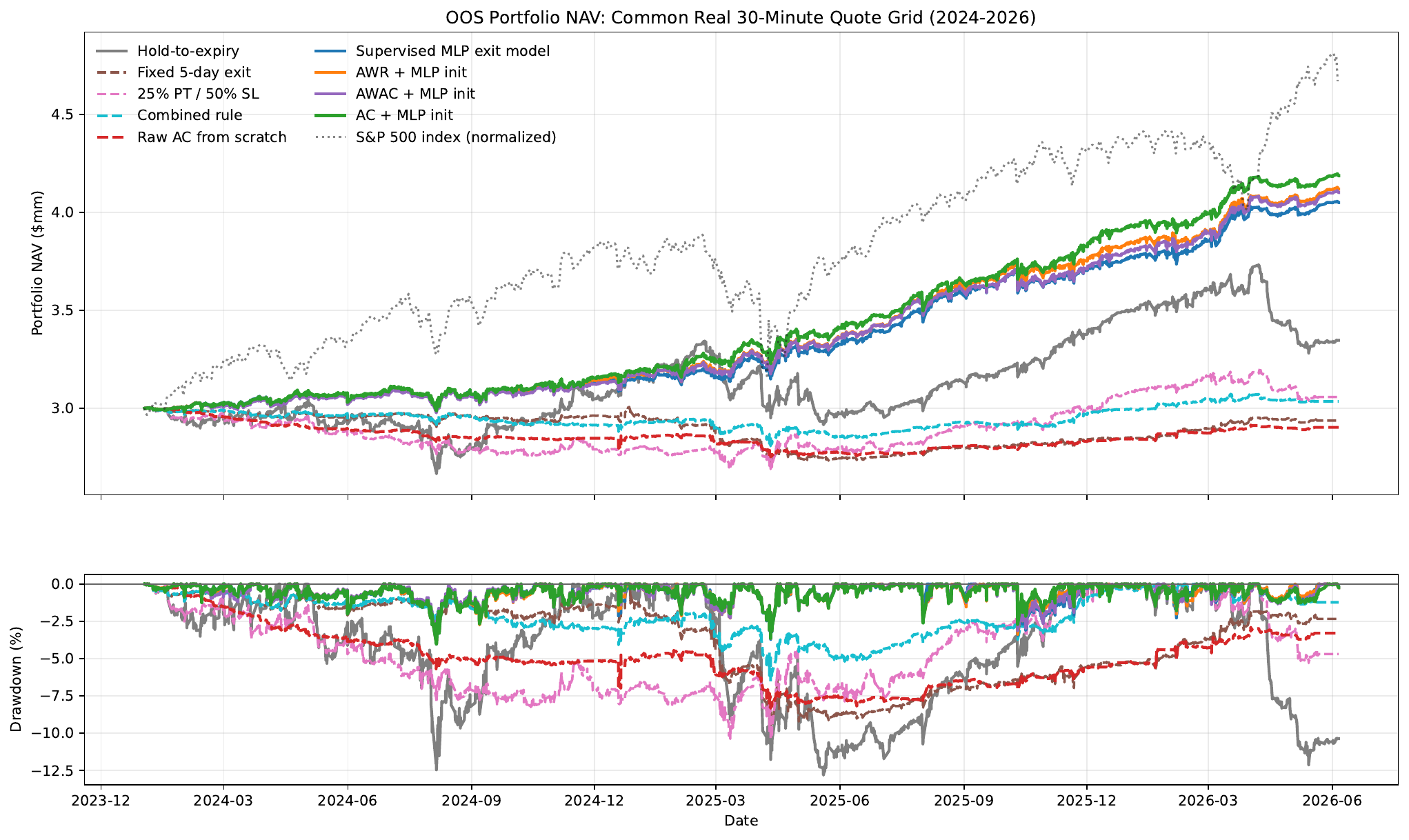}
  \caption{Out-of-sample portfolio NAV and drawdown for the deterministic strategies, representative reinforcement-learning runs, and the normalized S\&P 500 market reference over 2024--2026, after mapping each portfolio path to the common real 30-minute quote grid.}
  \Description{The figure compares portfolio NAV and drawdown paths for hold-to-expiry, three rule-based exits, the supervised MLP, representative raw Actor-Critic and MLP-initialized reinforcement-learning runs, and a separately identified normalized S\&P 500 market reference. These representative historical paths are not seven-seed mean curves. Multi-seed supervised and reinforcement-learning results are reported in Table~\ref{tab:rl_multiseed_results}.}
  \label{fig:main_nav_comparison}
\end{figure}

Figure~\ref{fig:main_nav_comparison} presents representative historical out-of-sample NAV and drawdown paths for these exit strategies. The normalized S\&P 500 series is shown separately in the figure as a market reference only; it is not treated as a short-straddle strategy or as a directly comparable exit-policy benchmark. The figure illustrates both final portfolio performance and the timing and severity of drawdowns.

\section{Results}
\label{sec:results}

The S\&P 500 reference series is constructed from daily closing prices of the S\&P 500 index downloaded from Yahoo Finance using the ticker \texttt{\string^GSPC} through the \texttt{yfinance} package \cite{yahooFinanceGSPC,yfinance}. It is included only as a normalized market reference and is not a short-straddle strategy.

\subsection{Portfolio Performance}

Table~\ref{tab:main_oos_results} reports the canonical out-of-sample performance panel for the passive and rule-based benchmark strategies. In addition to cumulative return, the table reports annualized return and risk measures, drawdown magnitude and duration, reward-to-risk ratios, hit rate, and trade count. This extended panel permits horizon-consistent comparison across strategies rather than comparing a multi-year cumulative return only with annualized Sharpe ratios. Because the passive and rule-based benchmark policies are deterministic, Table~\ref{tab:main_oos_results} reports their individual portfolio paths, while the supervised MLP and reinforcement-learning policies are summarized across independently trained seeds in Table~\ref{tab:rl_multiseed_results}. Reinforcement-learning results depend on stochastic optimization and are therefore reported separately as cross-seed means and standard deviations in Table~\ref{tab:rl_multiseed_results}. Figure~\ref{fig:main_nav_comparison} and Tables~\ref{tab:main_oos_results}--\ref{tab:rl_multiseed_results} report the main stress-budget specification; fixed-one-contract results are reported separately in the equal-notional robustness analysis.

Hold-to-expiry earns 11.56\% with a Sharpe ratio of 0.441 and MD of 12.82\%. The three mechanical exits fail to improve its return profile. Across seven independently trained runs, the supervised MLP achieves $35.57\%\pm1.41\%$ cumulative return, an average Sharpe ratio of $1.603\pm0.068$, and an average MD of $4.34\%\pm0.31\%$. All seven MLP runs earn between 33.91\% and 37.61\%, well above the 11.56\% hold-to-expiry result, indicating that the supervised improvement is not driven by a single network initialization. Its advantage over all three rules also indicates that it learns more than a constant holding horizon or a fixed profit-and-loss threshold.

\begin{table*}[t]
  \centering
  \caption{Canonical out-of-sample performance metrics for the deterministic benchmark exit policies over 2024--2026.}
  \label{tab:main_oos_results}
  \resizebox{\textwidth}{!}{%
  \begin{tabular}{@{}lrrrrrrrrrrrr@{}}
    \toprule
    Strategy & Cum. return & ARC & ASD & MD & MLD & IR$^*$ & IR$^{**}$ & IR$^{***}$ & Sharpe & Calmar & Hit rate & Trades \\
    \midrule
    Hold-to-expiry
      & 11.56\% & 4.62\% & 11.95\% & 12.82\% & 0.73
      & 0.386 & 0.360 & 0.063 & 0.441 & 0.360 & 56.9\% & 425 \\
    Fixed 5-day
      & -2.12\% & -0.88\% & 5.08\% & 9.78\% & 1.44
      & -0.173 & -0.090 & -0.006 & -0.150 & -0.090 & 57.4\% & 479 \\
    25\% PT / 50\% SL
      & 1.91\% & 0.78\% & 7.33\% & 10.45\% & 1.88
      & 0.107 & 0.075 & 0.004 & 0.144 & 0.075 & 66.5\% & 477 \\
    Combined rule
      & 1.15\% & 0.47\% & 4.31\% & 6.52\% & 2.04
      & 0.110 & 0.073 & 0.002 & 0.133 & 0.073 & 56.4\% & 495 \\
    \bottomrule
  \end{tabular}%
  }

  \begin{minipage}{\textwidth}
  \footnotesize
  \textit{Notes:} Return is cumulative; ARC, ASD, and MD are reported in percentage terms; MLD is reported in years. All NAV-based metrics are computed on the common real 30-minute quote grid over the shared OOS calendar window, with $T \approx 2.42$ years used for ARC. IR$^*$, IR$^{**}$, and IR$^{***}$ follow the definitions in the Evaluation Metrics subsection. Under these definitions, IR$^{**}$ is numerically equal to Calmar. Hit rate is the percentage of entered straddles with positive net episode PnL, and Trades is the number of entered straddles. Supervised and reinforcement-learning policies are reported separately in Table~\ref{tab:rl_multiseed_results} across independently trained seeds.
  \end{minipage}
\end{table*}

\begin{table*}[t]
  \centering
  \caption{Multi-seed canonical out-of-sample performance of the supervised and reinforcement-learning policies over 2024--2026. Values are mean $\pm$ sample standard deviation across independently trained seeds.}
  \label{tab:rl_multiseed_results}
  \textit{Panel A: Return, risk, and trading activity}\\[3pt]
  \resizebox{\textwidth}{!}{%
  \begin{tabular}{@{}lrrrrrrr@{}}
    \toprule
    Policy & Seeds & Cum. return & ARC & ASD & MD & Sharpe & Trades \\
    \midrule
    Supervised MLP & 7 & $35.57\%\pm1.41\%$ & $13.38\%\pm0.49\%$ & $8.12\%\pm0.22\%$ & $4.34\%\pm0.31\%$ & $1.603\pm0.068$ & $558.9\pm11.2$ \\
    Raw AC & 5 & $0.18\%\pm4.35\%$ & $0.06\%\pm1.77\%$ & $4.84\%\pm3.71\%$ & $8.70\%\pm2.67\%$ & $-0.102\pm0.339$ & $454.8\pm34.9$ \\
    AWR + MLP & 5 & $35.23\%\pm1.93\%$ & $13.26\%\pm0.67\%$ & $7.81\%\pm0.23\%$ & $3.95\%\pm0.11\%$ & $1.648\pm0.045$ & $561.4\pm12.5$ \\
    AWAC + MLP & 5 & $35.19\%\pm1.76\%$ & $13.25\%\pm0.61\%$ & $7.77\%\pm0.11\%$ & $3.88\%\pm0.05\%$ & $1.655\pm0.053$ & $557.0\pm9.0$ \\
    AC + MLP & 7 & $37.54\%\pm0.92\%$ & $14.15\%\pm0.34\%$ & $7.96\%\pm0.15\%$ & $3.96\%\pm0.16\%$ & $1.702\pm0.028$ & $567.3\pm6.2$ \\
    \bottomrule
  \end{tabular}%
  }

  \vspace{4pt}
  \textit{Panel B: Duration and additional performance measures}\\[3pt]
  \resizebox{\textwidth}{!}{%
  \begin{tabular}{@{}lrrrrrr@{}}
    \toprule
    Policy & MLD & IR$^*$ & IR$^{**}$ & IR$^{***}$ & Calmar & Hit rate \\
    \midrule
    Supervised MLP & $0.207\pm0.061$ & $1.649\pm0.075$ & $3.104\pm0.316$ & $0.693\pm0.182$ & $3.104\pm0.316$ & $55.8\%\pm1.4\%$ \\
    Raw AC & $2.022\pm0.744$ & $-0.124\pm0.323$ & $-0.012\pm0.162$ & $0.006\pm0.020$ & $-0.012\pm0.162$ & $52.3\%\pm2.8\%$ \\
    AWR + MLP & $0.272\pm0.041$ & $1.697\pm0.051$ & $3.359\pm0.163$ & $0.496\pm0.079$ & $3.359\pm0.163$ & $56.6\%\pm0.3\%$ \\
    AWAC + MLP & $0.282\pm0.024$ & $1.705\pm0.060$ & $3.411\pm0.142$ & $0.473\pm0.057$ & $3.411\pm0.142$ & $56.4\%\pm0.5\%$ \\
    AC + MLP & $0.283\pm0.029$ & $1.777\pm0.032$ & $3.573\pm0.119$ & $0.506\pm0.062$ & $3.573\pm0.119$ & $56.7\%\pm0.3\%$ \\
    \bottomrule
  \end{tabular}%
  }

  \vspace{2pt}
  \begin{minipage}{\textwidth}
  \footnotesize
  \textit{Notes:} Each canonical metric is first computed separately from the complete portfolio path generated by each independently trained policy and is then summarized across seeds. AC+MLP retains the archived per-run evaluation statistics used in the conference camera-ready. Other path panels use the common real 30-minute quote grid. The shared calendar test window does not imply identical NAV timestamp coverage in the archived AC+MLP summaries. Supervised MLP and AC+MLP each use seven seeds with labels $\{17,19,21,23,25,27,29\}$. The MLP models are independently trained, whereas the seven AC+MLP runs fine-tune from one common supervised checkpoint. They are not paired end-to-end training seeds, and AC+MLP standard deviations exclude variation across supervised checkpoints. Raw AC, AWR+MLP, and AWAC+MLP use five independent runs. ARC, ASD, and MD are reported in percentage terms, and MLD is reported in years. IR$^*$, IR$^{**}$, and IR$^{***}$ follow the definitions in the Evaluation Metrics subsection. Under these definitions, IR$^{**}$ is numerically equal to Calmar. Hit rate is the percentage of entered straddles with positive net episode PnL, and Trades is the number of entered straddles.
  \end{minipage}
\end{table*}

\subsection{Oracle Ceiling and the Remaining Imitation Gap}

The ex-post best-exit oracle observes the complete future path of each episode and is therefore a non-tradable diagnostic ceiling. Under the same portfolio simulator, common real 30-minute quote grid, and 26-bar minimum holding period, it achieves 373.02\% cumulative return, a Sharpe ratio of 6.305, and mean episode PnL of \$19,261. The corresponding mean PnL values are $\$1{,}830\pm\$73$ for the supervised MLP and approximately $\$1{,}932$ for AC+MLP, showing that the learned policies recover only a limited fraction of the ex-post opportunity available in the realized paths. This gap reflects both the oracle's perfect foresight and the limited predictability of future option-price paths from contemporaneous states; the oracle result is therefore not an attainable tradable target.

\subsection{Effect of MLP-Initialized Reinforcement Learning}

Table~\ref{tab:rl_multiseed_results} reports the cross-seed performance of the supervised and reinforcement-learning policies. Across seven independent training seeds, AC+MLP achieves an average cumulative return of $37.54\%\pm0.92\%$, an average ARC of $14.15\%\pm0.34\%$, and an average Sharpe ratio of $1.702\pm0.028$. Maximum drawdown remains stable at $3.96\%\pm0.16\%$, compared with $4.34\%\pm0.31\%$ for the supervised MLP. Mean episode PnL is approximately \$1,830 for MLP and \$1,932 for AC+MLP.

The seven-run AC+MLP result therefore increases average cumulative return by 1.97 percentage points and average Sharpe ratio by 0.099 relative to the supervised MLP. After filtering each run to actually entered episodes and averaging each policy's PnL across its available runs, the 583 matched entry episodes give an average AC+MLP difference of \$74 per episode. The paired $t$-test yields $t=0.993$ and $p=0.321$, while the Wilcoxon signed-rank test yields $p=0.027$. Because the two tests provide mixed evidence and entry episodes overlap through the 20-slot portfolio structure, the incremental RL improvement is interpreted as economically positive but not robustly statistically significant. These tests are descriptive matched-episode evidence rather than independent-trade inference. The 583 identifiers are matched across available entered runs; they are not episodes entered by every seed. Conditioning on entry also excludes the effect of skipped candidate opportunities. 

The learned policies also exit materially earlier than passive holding. Hold-to-expiry positions remain open for an average of 237 bars (median 256), whereas the supervised MLP averages 136.5 bars (median 138) and AC+MLP averages 136.8 bars (median 140) across their seven runs. At 13 intraday bars per trading day, the learned policies therefore reduce the average holding period from approximately 18.2 to 10.5 trading days. The similar aggregate holding periods of MLP and AC+MLP indicate that fine-tuning adjusts episode-specific exit decisions rather than merely learning a different constant holding horizon.

The alternative advantage-weighted updates provide supporting evidence on the role of supervised initialization. Across five seeds, AWR+MLP achieves a cumulative return of $35.23\%\pm1.93\%$ and a Sharpe ratio of $1.648\pm0.045$, while AWAC+MLP achieves $35.19\%\pm1.76\%$ and $1.655\pm0.053$. These variants broadly preserve the supervised MLP's performance but do not provide the same average incremental improvement as Actor--Critic. They are therefore interpreted as robustness variants of the supervised-first framework rather than as the principal performing models.

\subsection{Equal-Notional Robustness}

Under fixed-one-contract sizing, the average ordering of AC+MLP, the supervised MLP, and hold-to-expiry is preserved, although the AC+MLP increment narrows to 0.45 percentage points versus a single supervised checkpoint and is not statistically significant ($p=0.645$). This robustness test retains all matched candidate records, including zero-PnL non-entry rows, whereas the main updated paired comparison conditions on actual entry; their p-values therefore use different samples. Dynamic sizing therefore amplifies but does not solely generate the average ranking; complete common-grid results appear in Table~\ref{tab:equal_notional_results} in the appendix.

\subsection{Reward-Shaping Ablation}

The reward ablation is a separate within-ablation comparison, not the principal seven-seed performance batch. To test whether the Actor--Critic ranking is an artifact of the hand-specified reward coefficients, we repeat the AC+MLP experiment across the common seed set $\{1,3,5,7,11\}$ under four reward specifications. Table~\ref{tab:reward_ablation} reports the results. The full reward achieves a cumulative return of $35.56\%\pm2.73\%$ and a Sharpe ratio of $1.661\pm0.093$. Removing the CVaR term produces $36.80\%\pm1.07\%$ and $1.677\pm0.057$, while removing the VRP component produces $34.30\%\pm1.87\%$ and $1.655\pm0.066$. The direct-reward specification achieves $35.11\%\pm1.18\%$ and $1.627\pm0.053$.

The four specifications produce broadly similar out-of-sample performance, so the effectiveness of MLP initialization is not driven by one reward-shaping component. The CVaR term does not generate the return ranking: removing it slightly raises average return and Sharpe but worsens average maximum drawdown from 4.00\% to 4.16\%. Removing the VRP term reduces average return but improves average maximum drawdown to 3.70\%. These results suggest that the shaping components influence the return--drawdown trade-off rather than mechanically creating the overall advantage of the supervised-first policy.

\begin{table}[t]
  \centering
  \caption{Five-seed reward-shaping ablation for AC+MLP over 2024--2026. Values are mean $\pm$ sample standard deviation.}
  \label{tab:reward_ablation}
  \resizebox{\linewidth}{!}{%
  \begin{tabular}{@{}lrrrr@{}}
    \toprule
    Reward & Cum. return & Sharpe & MaxDD & Mean PnL \\
    \midrule
    Full shaped
      & $35.56\%\pm2.73\%$
      & $1.661\pm0.093$
      & $-4.00\%\pm0.03\%$
      & $\$1{,}830\pm\$140$ \\
    No CVaR
      & $36.80\%\pm1.07\%$
      & $1.677\pm0.057$
      & $-4.16\%\pm0.17\%$
      & $\$1{,}894\pm\$55$ \\
    No VRP
      & $34.30\%\pm1.87\%$
      & $1.655\pm0.066$
      & $-3.70\%\pm0.20\%$
      & $\$1{,}765\pm\$96$ \\
    Direct
      & $35.11\%\pm1.18\%$
      & $1.627\pm0.053$
      & $-3.93\%\pm0.14\%$
      & $\$1{,}807\pm\$61$ \\
    \bottomrule
  \end{tabular}%
  }

  \vspace{2pt}
  \begin{minipage}{\linewidth}
  \footnotesize
  \textit{Notes:} Every specification uses Seeds 1, 3, 5, 7, and 11. NAV-based metrics use the common real 30-minute quote grid and common calendar horizon; Mean PnL is computed from episode-level outcomes. ``Direct'' removes both the CVaR and VRP shaping terms while retaining realized trading performance and transaction costs.
  \end{minipage}
\end{table}

\subsection{Raw Reinforcement Learning From Scratch}

Actor--Critic trained from scratch performs poorly and is highly sensitive to stochastic training. Across five independent runs, raw AC achieves an average cumulative return of $0.18\%\pm4.35\%$, an average Sharpe ratio of $-0.102\pm0.339$, an average maximum drawdown of $8.70\%\pm2.67\%$, and an average mean episode PnL of only $\$10\pm\$226$. Across 564 matched entry episodes, its average PnL difference relative to the supervised MLP is $-\$1{,}233$. The difference is statistically significant under both the paired $t$-test ($t=-4.932$, $p=1.07\times10^{-6}$) and Wilcoxon signed-rank test ($p=4.55\times10^{-7}$). This contrast shows that reinforcement learning alone does not explain the performance of the two-stage framework.





\section{Limitations}

Best-quote execution abstracts from order latency, displayed depth, partial fills, volume-weighted execution prices, and size-dependent market impact. Realized slippage may therefore be larger, especially in volatile or near-expiry periods. The main seven-seed comparison is not an end-to-end paired-seed experiment because AC+MLP shares one supervised initialization.

The 2024--2026 out-of-sample period includes brief volatility spikes, notably in August 2024 and April 2025, but no sustained volatility shock comparable to 2008 or March 2020. The observed approximately 4\% drawdowns should therefore not be interpreted as upper bounds on naked short-straddle tail risk. The simulator includes bid--ask execution, margin constraints, maintenance liquidation, and risk-aware training, but it cannot fully reproduce crisis-period price gaps, volatility jumps, spread widening, liquidity shortages, or forced deleveraging. Evaluation over longer and more severe volatility regimes remains necessary.

The synthetic-fill sensitivity check is limited to the supervised MLP and one AC+MLP run. It establishes that these audited paths do not execute exits on synthetic bars, but it does not imply that synthetic continuation states are irrelevant to every possible policy or market regime.

The incremental AC+MLP advantage is smaller under fixed-one-contract sizing, indicating that part of its economic magnitude in the main stress-budget specification is associated with the timing of gains and subsequent slot-level capital scaling. Although the average policy ordering is preserved, the equal-notional episode-level difference between AC+MLP and the supervised MLP is not statistically significant. The reported RL increment should therefore be interpreted as an average economic improvement that is amplified by dynamic position sizing rather than as a uniform episode-level advantage.

\section{Conclusion}

This paper develops a two-stage framework for SPX short-straddle exit timing in which an oracle-guided supervised MLP initializes reinforcement-learning policies. The supervised policy substantially outperforms passive holding and representative mechanical exits. Across seven seeds, Actor--Critic fine-tuning provides a positive average economic improvement, while matched-episode tests provide mixed evidence and the increment is not robustly statistically significant. Raw Actor--Critic performs poorly, while the reward and equal-notional checks show that the main average ranking is not solely generated by one shaping component or dynamic position sizing. Overall, supervised initialization is what makes reward-based fine-tuning viable in this sparse and path-dependent trading problem.

\section{Code and Data Availability}

The code used for the main experiments is available at:
\url{https://github.com/Justinnnn0313/spx-short-straddle-rl}.

The underlying SPX option quotations are proprietary Cboe data accessed through an institutional license held by one of the co-authors. The license does not permit redistribution of the source option-quote files, and these data are therefore not included in the public repository. Researchers with lawful access to equivalent Cboe SPX option data can use the released code to reproduce the data-processing, model-training, and portfolio-evaluation pipeline.


\begin{acks}
The authors gratefully acknowledge financial support from the Odyssey Research Programme at the School of Physical and Mathematical Sciences, Nanyang Technological University.
\end{acks}

\section*{Ethics and Privacy Statement}

This study uses licensed financial-market data and involves no human subjects or personally identifiable information. The results are intended for research and should not be interpreted as investment advice or as evidence that the policies eliminate short-volatility tail risk.
\bibliographystyle{ACM-Reference-Format}
\bibliography{references}

\appendix

\section{Implementation and Portfolio Construction Details}
\label{app:implementation_details}

This appendix reports the implementation details used for the final reinforcement-learning experiments and the out-of-sample portfolio simulation. The reported reinforcement-learning configurations were fixed after preliminary pilot runs and then held unchanged for the final 2024--2026 out-of-sample evaluation. No model hyperparameters were selected using the 2024--2026 out-of-sample test period.

\subsection{Environment and State Variables}

The trading environment uses an exit-timing formulation for SPX short-straddle positions. At each decision time, the policy observes a vector of market, quote, and position-state variables. In the final reinforcement-learning runs, the observation setting uses the position-augmented state representation. The base state contains time-to-maturity, time-of-day, variance-risk-premium, relative bid--ask spread, model score, recent realized-volatility change, quote-availability indicators, synthetic-bar indicators, simulated time-value information, current position, and normalized running PnL. The position-augmented representation adds holding time, entry maturity, remaining maturity relative to entry maturity, current exit value relative to entry value, underlying return since entry, and moneyness drift since entry.

All observations are numerically cleaned and clipped to the interval $[-20,20]$. The final position-augmented state has dimension 17. Invalid actions are remapped to the hold action. Exits are permitted only when the quote is exitable and, except at terminal settlement, after the minimum holding period has elapsed.

\subsection{Synthetic-Quote Audit and Fill Sensitivity}
\label{app:synthetic_fill_audit}

Synthetic continuation bars are used when a complete real-time call--put quotation is unavailable but an episode must continue toward its exit or expiration. These bars are explicitly identified by the synthetic-bar state indicator and are not treated as observed Cboe quotations. Across the complete episode set, 301,645 of 5,585,856 bars are synthetic, corresponding to 5.40\% of all observations. The share is 5.69\% during 2016--2023 and 5.03\% during the 2024--2026 out-of-sample period. At the episode level, 20.56\% of out-of-sample episodes contain at least one synthetic continuation bar, although such episodes may contain predominantly real quotations.

We further examine whether learned-policy performance depends on optimistic synthetic exits. In the single-checkpoint supervised MLP audit, none of the 559 executed exits occurs on a synthetic bar. In a single AC+MLP run used for the fill-sensitivity audit, none of the 562 executed exits occurs on a synthetic bar. We then multiply the synthetic half-spread assumption by 1.25, 1.50, and 2.00 while leaving real-quote execution unchanged. Because neither policy executes an exit on a synthetic bar, the conservative synthetic-fill assumptions do not change their realized portfolio results.

\begin{table}[t]
  \centering
  \caption{Synthetic-bar coverage and fill-sensitivity audit.}
  \label{tab:synthetic_fill_audit}
  \resizebox{\linewidth}{!}{%
  \begin{tabular}{@{}lrrr@{}}
    \toprule
    Period & Synthetic bars & Total bars & Share \\
    \midrule
    Full sample
      & 301,645 & 5,585,856 & 5.40\% \\
    Training, 2016--2023
      & 178,791 & 3,144,059 & 5.69\% \\
    OOS, 2024--2026
      & 122,854 & 2,441,797 & 5.03\% \\
    \bottomrule
  \end{tabular}%
  }

  \vspace{3pt}
  \begin{minipage}{\linewidth}
  \footnotesize
  \textit{Notes:} In the OOS sample, 2,014 of 9,794 episodes (20.56\%) contain at least one synthetic continuation bar. The supervised MLP has 0 synthetic exits among 559 exits, while the AC+MLP sensitivity run has 0 among 562 exits.
  \end{minipage}
\end{table}

Under both the baseline and doubled synthetic half-spread assumptions, the supervised MLP retains a cumulative return of 35.01\%, a Sharpe ratio of 1.651, and a maximum drawdown of 4.04\%. That AC+MLP run retains a cumulative return of 39.58\%, a Sharpe ratio of 1.802, and a maximum drawdown of 4.02\%; these are historical single-run audit values, not the main multi-seed estimates in Table~\ref{tab:rl_multiseed_results}. This invariance does not imply that synthetic bars are economically irrelevant in every application; rather, it shows that the reported fills of these audited learned-policy paths are not generated by exits executed on synthetic quotations.

\subsection{Reward Parameters}

The final RL experiments use the shaped reward specification described in the main text. The reward parameters are held fixed across the RL comparisons:
\[
  \alpha = 1.5,\qquad
  \beta_{\mathrm{VRP}} = 0.5,\qquad
  \lambda = 0.2,\qquad
  \mu = 0.05.
\]
The minimum holding period is 26 intraday bars. With 13 intraday bars per trading day, this corresponds to approximately two trading days.

\subsection{Reinforcement-Learning Hyperparameters}

Table~\ref{tab:appendix_rl_hyperparameters} summarizes the training configurations used for the reported RL policies. The MLP-initialized models load the supervised exit model into the actor before RL fine-tuning, while the raw Actor-Critic baseline uses random initialization.

\begin{table*}
  \centering
  \caption{Reinforcement-learning configurations used in the final reported experiments.}
  \label{tab:appendix_rl_hyperparameters}
  \resizebox{\linewidth}{!}{%
  \begin{tabular}{@{}lrrrrrrrrr@{}}
    \toprule
    Model & Init. & Iter. & LR & $\gamma$ & Episodes/update & Entropy coeff. & BC coeff. & Value coeff. & Runs \\
    \midrule
    Raw Actor-Critic
      & Random & 300 & $5{\times}10^{-6}$ & 0.997 & 8
      & 0.0010 & 0.00 & 0.5 & 5 \\
    Actor-Critic + MLP
      & MLP & 300 & $5{\times}10^{-6}$ & 0.997 & 8
      & 0.0010 & 0.05 & 0.5 & 7 \\
    AWR + MLP
      & MLP & 120 & $1{\times}10^{-5}$ & 0.997 & 32
      & 0.0005 & 0.01 & 0.5 & 5 \\
    AWAC + MLP
      & MLP & 120 & $1{\times}10^{-5}$ & 0.997 & 32
      & 0.0005 & 0.01 & 0.5 & 5 \\
    \bottomrule
  \end{tabular}%
  }
\end{table*}

\begin{table*}[t]
  \centering
  \caption{Equal-notional out-of-sample robustness over 2024--2026.}
  \label{tab:equal_notional_results}
  \resizebox{\linewidth}{!}{%
  \begin{tabular}{@{}lrrrrrr@{}}
    \toprule
    Policy & Return & Sharpe & MaxDD & Total PnL & Mean PnL & Trades \\
    \midrule
    Hold-to-expiry
      & 15.67\% & 0.544 & -12.63\%
      & \$470,077 & \$818 & 454 \\
    Supervised MLP
      & 35.04\% & 1.566 & -3.94\%
      & \$1,051,183 & \$1,803 & 577 \\
    AC + MLP
      & $35.49\%\pm1.93\%$
      & $1.590\pm0.056$
      & $-4.00\%\pm0.12\%$
      & $\$1{,}064{,}626\pm\$57{,}979$
      & $\$1{,}826\pm\$99$
      & $574.6\pm8.1$ \\
    \bottomrule
  \end{tabular}%
  }

  \vspace{2pt}
  \begin{minipage}{\linewidth}
  \footnotesize
  \textit{Notes:} Every entered position contains exactly one short-straddle contract. AC+MLP values are mean $\pm$ sample standard deviation across seven independently trained policies. NAV-based metrics use the common real 30-minute quote grid and common calendar horizon. The MLP row uses one supervised checkpoint, rather than the seven-model baseline in the main panel. All other execution, margin, and portfolio assumptions are unchanged.
  \end{minipage}
\end{table*}

The principal AC+MLP runs use seeds 17, 19, 21, 23, 25, 27, and 29, with a common supervised initialization. Raw Actor--Critic, AWR+MLP, and AWAC+MLP are each summarized across five independent runs. Separately, the reward ablation uses the matched seed set $\{1,3,5,7,11\}$ for every reward specification so that differences are attributable to reward design rather than stochastic initialization.

For the AWR and AWAC variants, the advantage-weighting temperature is set to 2.0, the maximum importance weight is clipped at 20, the minibatch size is 512, and one update epoch is used per training iteration. All RL models are trained on 2016--2023 episodes and evaluated out of sample on 2024--2026 episodes.

\subsection{Out-of-Sample Positioning and Capital Allocation}

The portfolio simulator starts with total capital
\[
  W_0 = \$3{,}000{,}000
\]
and divides it into 20 capital slots. Each slot therefore starts with
\[
  W_0^{\mathrm{slot}} = \frac{W_0}{20}.
\]
Each slot can hold at most one straddle episode at a time. When a slot finishes an episode, its updated capital is recycled into the next available eligible episode. The portfolio NAV is the sum of the mark-to-market NAV across all slots.

For each out-of-sample trading date, the simulator selects one candidate straddle episode. Eligible starts have 18--22 trading days to maturity and moneyness in $[0.975,1.025]$. Candidate episodes are sorted by closeness to the target maturity, closeness to at-the-money moneyness, smaller relative spread, and earlier quote time. The target maturity is 20 trading days. This rule produces a daily sequence of tradable candidate episodes without using future out-of-sample performance information.

Entry is attempted at the selected episode's first quote. If capital or margin constraints prevent one-contract entry, the evaluator defers entry and retries at subsequent real quotes. Each portfolio slot maintains its own marked-to-market NAV. When a position closes, its realized net PnL remains in that slot and becomes the capital base for the next position assigned to it. The main stress-budget specification is therefore dynamically compounded at the slot level rather than rebalanced to a fixed notional after each episode.

Position size is determined by a stress-budget rule. Let $c=\$1.50$ denote the commission per option leg, per contract, and per transaction event, consistent with the definition in the main text. For a short straddle entered at premium $P_t^{\mathrm{entry}}$ with underlying level $S_t$, the per-contract stress loss is
\[
  L_t^{\mathrm{stress}}
  =
  \max\{0,\; 0.15 S_t - P_t^{\mathrm{entry}}\}M
  +
  4c.
\]
Here $M=100$ is the option contract multiplier. The term $4c=\$6.00$ is the full round-trip commission for one straddle contract: two option legs at entry and two option legs at exit.
\[
  n_t
  =
  \left\lfloor
  \frac{0.50 W_t^{\mathrm{slot}}}{L_t^{\mathrm{stress}}}
  \right\rfloor,
\]
subject to available-capital and margin constraints.

For the equal-notional robustness specification, the stress-budget sizing rule is replaced by
\[
n_t^{\mathrm{equal}}=1.
\]
For every entered episode, this removes variation in contract count caused by prior slot performance while retaining the same candidate episodes, exit policies, bid--ask execution, commissions, and portfolio-slot mechanics.

The number of entered trades can differ from the stress-budget specification because fixing the contract count changes whether an otherwise identical candidate passes the available-capital and margin constraints.

For short straddles, the required entry capital per contract is
\[
  K_t
  =
  \max\{P_t^{\mathrm{entry}}M,\; 0.20 S_t M\}.
\]
The final contract count is capped so that the slot has enough capital to cover required entry capital plus the one-way entry commission. During the episode, the maintenance requirement is
\[
  K_t^{\mathrm{maint}}
  =
  0.15 S_t M n_t.
\]
If the mark-to-market slot NAV falls below this requirement, the simulator liquidates the position at the current exit fill and records the liquidation in the portfolio accounting.

\subsection{Execution and Accounting}

For a short straddle, entry is executed at the observed straddle bid and discretionary exit is executed at the observed straddle ask. If the position reaches expiration, settlement value is computed as
\[
  V_T^{\mathrm{settle}}
  =
  |S_T-K|,
\]
Here, $K$ is the strike. Dollar PnL for a short straddle with $n_t$ contracts is computed as
\[
  \mathrm{PnL}
  =
  \left(P_t^{\mathrm{entry}} - P_t^{\mathrm{exit}}\right)Mn_t
  -
  \mathrm{Commission},
\]
For a discretionary round-trip exit,
\[
\mathrm{Commission}=4c n_t,
\]
whereas each individual entry or exit event incurs $2c n_t$. The same total round-trip commission applies when the second transaction event is terminal settlement.

Analogous accounting applies when the position expires and the exit value is replaced by the settlement value. Let
\[
r_t=\frac{W_t}{W_{t-1}}-1
\]
denote the 30-minute simple return of aggregate marked-to-market portfolio NAV. The reported Sharpe ratio is
\[
\mathrm{Sharpe}
=
\frac{\bar r}{s(r)}
\sqrt{3276},
\]
Here, $\bar r$ and $s(r)$ are the sample mean and sample standard deviation of the 30-minute portfolio-return series. Returns are not adjusted for a separate risk-free rate; equivalently, the calculation assumes a zero risk-free return at the 30-minute horizon. The same convention is applied to all strategies.

Portfolio metrics are computed from marked-to-market NAV; common-grid normalization applies except for the archived AC+MLP evaluation statistics retained in Table~\ref{tab:rl_multiseed_results}. MLD is measured as the longest elapsed calendar time, expressed in years, during which portfolio NAV remains below its previous running maximum. Hit rate is the fraction of entered straddles with positive net episode PnL, and trade count is the number of entered straddles.

\end{document}